\documentclass[
    superscriptaddress,
    amssymb,
    amsmath,
    aps,
    prl,
    twocolumn,
    nofootinbib
]{revtex4-1}

\usepackage{hyperref}
\usepackage{amsmath}
\usepackage{amssymb}
\usepackage{graphicx}
\usepackage{bm}
\usepackage[usenames,dvipsnames,svgnames,table]{xcolor}
\usepackage{aas_macros}
\usepackage{orcidlink}
\usepackage[normalem]{ulem}

\newcommand{\e}{\operatorname{e}}

\usepackage[labelfont=bf]{caption}
\hypersetup{
    colorlinks=true,
    linkcolor=Blue,
    filecolor=Blue,
    urlcolor=MidnightBlue,
    citecolor=Blue,
   pdftitle={Cooling age}
}

\begin{document}

%\preprint{}

\title{Scalarized White Dwarfs: Superdense Without Strangeness}


\author{Sof\'ia Vidal\orcidlink{0000-0003-0143-0427}}
\thanks{Corresponding author}
\email{sofia.vidal@ut.ee}
\affiliation{Laboratory of Theoretical Physics, Institute of Physics, University of Tartu,
W. Ostwaldi 1, 50411 Tartu, Estonia
}

\author{Aneta Wojnar\orcidlink{0000-0002-1545-1483}}
\email[E-mail: ]{aneta.wojnar@uwr.edu.pl}
\affiliation{Institute of Theoretical Physics, University of Wroc\l aw, plac Maxa Borna 9, 50-206 Wroc\l aw, Poland}

\author{Laur J\"arv\orcidlink{0000-0001-8879-3890}}
\email[E-mail: ]{laur.jarv@ut.ee}
\affiliation{Laboratory of Theoretical Physics, Institute of Physics, University of Tartu,
W. Ostwaldi 1, 50411 Tartu, Estonia
}

\begin{abstract}
Additional families of compact stellar objects are commonly associated with strange stars composed of exotic high-density matter, or phase transitions. We show that a qualitatively similar phenomenon can arise from gravity alone. In massive scalar-tensor gravity, spontaneous scalarization of white dwarfs generates an additional branch of compact equilibrium configurations while the matter sector is described by the standard Chandrasekhar equation of state. The resulting mass-radius relation resembles that of strange-dwarf and third-family sequences, despite the absence of exotic matter or a phase transition. We derive an analytical estimate for the onset of white-dwarf scalarization, identify parameter regions compatible with current binary-pulsar and cosmological considerations and numerically obtain \textit{scalarized dwarf} solutions within these parameter regions. The predicted configurations, with subsolar masses and radii of only a few thousand kilometers, provide a distinctive target for searches for anomalous compact objects. 
\end{abstract}

\maketitle

\textit{Introduction.} 
The appearance of new families of compact stars is usually associated with nontrivial high-density microphysics. Deconfined quark matter can give rise to strange and hybrid stars \cite{Ivanenko:1965dg,Itoh:1970uw,Bodmer:1971we,Witten:1984rs,Farhi:1984qu}, while sufficiently strong phase transitions may produce disconnected sequences and third-family configurations \cite{Alvarez-Castillo:2018pve,Benic:2014jia,Alvarez-Castillo:2017qki}. Although deconfined quark matter has been experimentally produced in the form of quark-gluon plasma, its existence in stellar interiors remains hypothetical, and no unambiguous observational evidence for twin stars \cite{Christian:2017jni} has yet been found. Current observations, including neutron stars with masses exceeding $2M_\odot$, place important constraints on the equation of state (EoS) but do not uniquely establish the presence of deconfined quark matter, as they may also be explained by suitable hadronic models or modifications of gravity \cite{Olmo:2019flu,Lope-Oter:2024egz}.

A particularly intriguing example is provided by strange dwarfs: configurations containing a compact strange-matter core surrounded by ordinary matter, whose masses and radii can nevertheless remain comparable to those of white dwarfs \cite{Glendenning:1994zb,Glendenning:1994sp,Vartanyan:2004dy}. Such objects illustrate how exotic microscopic physics can generate stellar sequences qualitatively different from the standard white-dwarf sequence.

Extensions of general relativity (GR), however, can provide another source for nonstandard astrophysical objects. A particularly striking nonlinear phenomenon in theories with a scalar field nonminimally coupled to curvature is spontaneous scalarization, in which the GR solution becomes unstable to the growth of a nontrivial scalar field, while an alternative solution branch emerges for strong gravity and superdense environment. Originally envisioned for neutron stars \cite{Damour:1993hw}, the scalarization phenomenon has been also extensively explored for black holes  \cite{Doneva:2017bvd,Silva:2017uqg,Herdeiro:2020wei,Zhang:2021nnn}, see Ref.\ \cite{Doneva:2022ewd} for an overview. On the other hand, the effects of the scalar field on white-dwarf structure, mass-radius relation, crystallization, and thermal evolution have recently seen thorough investigations \cite{Benvenuto:1999dk,Koyama:2015oma,Babichev:2016jom,Jain:2015edg,Saltas:2018mxc,Wojnar:2020ckw,Alam:2023ojf,Vidal:2024wto,Bachs-Esteban:2024vym,Vidal:2026dlf}, but only as a modification of the conventional white-dwarf sequence, reviewed in Ref.\ \cite{Olmo:2019flu}. A common expectation has been that white dwarfs are not compact enough to trigger spontaneous scalarization, because the required stronger nonlinear coupling to gravity would violate observational constraints.

In this Letter, we argue that scalarization is possible. If the nonminimal scalar field is also massive, the observational limits are considerably weaker and depend on scale \cite{Perivolaropoulos:2009ak,Hohmann:2013rba,Ramazanoglu:2016kul,Seymour:2019tir,Tuna:2022qqr,Kuan:2023hrh} which opens an interesting window of opportunity. We show that using the standard Chandrasekhar EoS without introducing exotic matter or a phase transition, sufficiently strong scalarization generates an additional family of compact equilibrium configurations. The resulting mass-radius and mass-central-density relations exhibit qualitatively distinct structures from those associated with strange dwarfs and third-family configurations, including the emergence of new sequences at lower masses. This suggests a genuinely different family of compact objects, arising from gravitational rather than exotic-matter effects. We derive an analytical estimate for the onset of white-dwarf scalarization and show that a parameter region remains compatible with binary-pulsar and cosmological considerations.

 Our results also provide a new motivation to search for anomalously compact white-dwarf-like objects, for which independent mass and radius measurements may distinguish the scalarized sequence from the standard Chandrasekhar sequence.

\textit{Scalarized solutions.} We consider {a massive} Damour-Esposito-Far\`ese (DEF) scalar-tensor theory in Einstein frame \cite{Damour:1993hw},
\begin{equation}
    S=\frac{1}{16\pi}\int d^4x\sqrt{-g}
    \left[
    R-2\partial_\mu\varphi\,\partial^\mu\varphi
    -V(\varphi)
    \right]
    +S_m\;,
    \label{eq:action}
\end{equation}
where the action $S_m=S_m\!\left[A^2(\varphi)g_{\mu\nu},\chi\right]$ describes how the matter fields $\chi$ couple nonminimally via the function $A(\varphi)=e^{\beta_0\varphi^2/2}$ 
with constant theory parameter $\beta_0$, to a scalar field with mass $m_\varphi$ as encoded in the potential $V(\varphi)=2m_\varphi^2\varphi^2$. For this choice, $\varphi=0$ defines the GR {solution}. Linear scalar
perturbations around GR obey
\begin{equation}
    \Box_{\rm GR}\delta\varphi
    =
    m_{\rm eff}^2\,\delta\varphi,
    \qquad
    m_{\rm eff}^2
    =
    m_\varphi^2-4\pi\beta_0\tilde T\; ,
    \label{eq:scalarfieldperturbation}
\end{equation}
where $\tilde T=3\tilde p-\tilde\rho$ is the trace of the matter
energy-momentum tensor in the physical Jordan frame. For $\beta_0<0$, the matter contribution can
render $m_{\rm eff}^2<0$, allowing for a tachyonic instability of the
GR solution. The onset of scalarization is determined by the marginal,
zero-frequency mode.

A simple estimate of this threshold follows by treating the white dwarf
in the weak-field, constant-density approximation,
$\tilde T\simeq-\tilde\rho_0$, with 
$\tilde\rho_0 = 3M/4\pi R^3$ being the average density, defined through the total stellar mass $M$ and the stellar radius $R$.
Using the ansatz ${\delta\varphi = \e^{-i\omega_n t}Y_{lm}(\theta, \varphi)\frac{\psi_n(r)}{r}}$ for the field perturbation, the perturbation equation reduces to
\begin{equation}
    \frac{d^2\psi_n}{dr^2} + \left[\omega_n^2 -  m_{\rm eff}^2\right]\psi_n =  0 \;,
    \label{eq:perturbationequation_scalarfield}
\end{equation}
with eigenmodes $\omega_n$, {$n\in\mathbb{N}_0$}.
Imposing regularity at the origin and matching to the
exterior solution at $r=R$ in the massless case, $m_\varphi=0$, gives
\begin{equation}
m_{\rm eff} R = \frac{\pi}{2},
%\qquad
%m_{\rm eff, massless}^2=4\pi|\beta_0|\tilde\rho_0,
\end{equation}
and hence introducing compactness $C\equiv\frac{M}{R}$
\begin{equation}
|\beta_{\rm crit}|
=
\frac{\pi^2}{12}\frac{R}{M}
=
\frac{\pi^2}{12C}.
\end{equation}
Thus, the scalarization threshold scales inversely with stellar compactness, explaining why white dwarfs require substantially larger \(|\beta_0|\) than neutron stars. For a typical white dwarf with $M=0.6M_\odot$ and
$R\simeq1.4R_\oplus$, this gives
$|\beta_{\rm crit}|\sim10^4$, whereas close to the Chandrasekhar limit,
$M\simeq1.4M_\odot$ and $R\simeq10^3\,{\rm km}$, it decreases to
${|\beta_{\rm crit}|\sim4\times10^2}$.

For a massive scalar, the exterior field is Yukawa suppressed,
$\psi_{\rm ext}\propto e^{-m_\varphi r}$. Matching the interior and exterior solutions at the stellar surface
yields
\begin{equation}
    m_{\rm eff}\cot(m_{\rm eff}R)=-m_\varphi,
   \label{eq:massiveDEF_critbetacondition}
\end{equation}
which determines $\beta_{\rm crit}$ as a function of $m_\varphi$ and $C$. The scalar mass therefore competes with the
matter-induced tachyonic term and progressively suppresses
scalarization as its Compton wavelength becomes shorter than the
stellar scale.

\begin{figure}[ht]
    \centering
    \includegraphics[width=1.\linewidth]{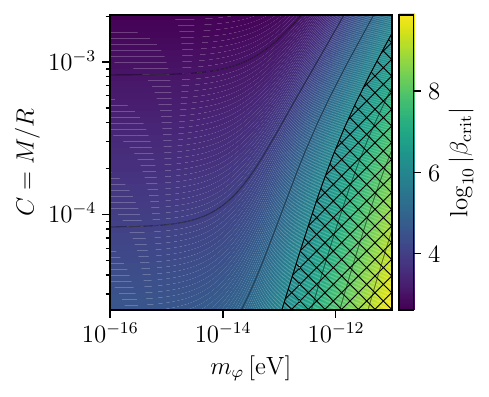}
    \caption{Absolute value of $\beta_{\rm crit}$, the threshold for scalarization, as a function of $m_\varphi$ and ${C=M/R}$. The critical parameter $\beta_\mathrm{crit}$ is obtained by assuming  ${\tilde\rho_0 = 3M/4\pi R^3}$ and solving \eqref{eq:massiveDEF_critbetacondition}. Contour lines range from ${|\beta_\mathrm{crit}| = 10^3}$ in the upper left to $10^{9}$ in the lower right corner. The hatched region, ${|\beta_\mathrm{crit}| > 10^6}$, indicates parameters for which scalarization may also be possible for main-sequence stars.}
    \label{fig:massiveDEF_critbeta}
\end{figure}

Using $M$ and $R$ from the corresponding GR stellar models, we solve this condition numerically, with the resulting thresholds shown in Fig.\ \ref{fig:massiveDEF_critbeta}. For the most compact white dwarfs, $|\beta_\mathrm{crit}|$ ranges from $\sim360$ for light scalar fields to $\sim3\times10^9$ for larger $m_\varphi$. For low-compactness stars, the corresponding thresholds are considerably higher, $3\times10^4\lesssim|\beta_\mathrm{crit}|\lesssim7\times10^{13}$.

\begin{table}[ht]
    \setlength{\tabcolsep}{5pt}
    \centering
    \begin{tabular}{ |c|c|c|c| } 
         \hline
         \rule{0pt}{2.5ex} $-\beta_0$ & $m_\varphi$ [$1$] & scalarized solution & branches \\ 
         [0.5ex]
         \hline
         \rule{0pt}{2.5ex} $1 \times 10^2$ & $10$ & no &  - \\
         $2 \times 10^2$ & $10$ & yes & $1$ \\
         $3 \times 10^2$ & $10$ & yes & $1$ \\
         $5 \times 10^2$ & $10$ & yes & $2$ \\
         $1 \times 10^3$ & $10-10^5$ & yes & $2$ \\
         $2 \times 10^3$ & $10-10^4$ & yes & $3$ \\
         $5 \times 10^3$ & $10$ & yes & {$4$} \\
         $1 \times 10^4$ & $10$ & yes & {$4$} \\ [0.5ex]
         \hline
    \end{tabular}
    \caption{
        Values of $\beta_0$ and {dimensionless scalar field mass \eqref{eq:scalarfieldmassdim}} for which numerical results have been calculated, whether scalarization takes place and number of branches found. Only branches that begin diverging from GR at $\rho_c \leq 10^{12}\, \rm g \,\rm {cm}^{-3}$ are included.
    }
    \label{table:numericalsolutions}
\end{table}

\begin{figure}[ht]
    \centering
    \includegraphics[width=1.\linewidth]{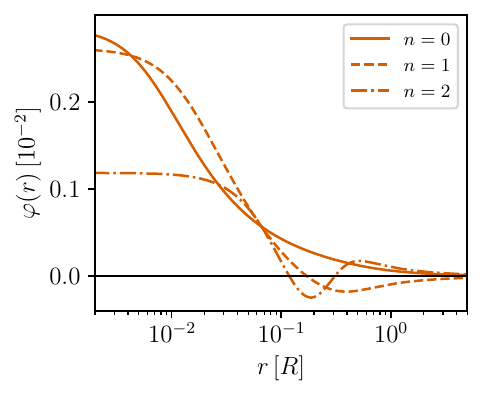}
    \caption{
        Radial scalar field profiles for a white dwarf with central density ${\rho_c = 7.7\times10^{10}\, \rm g \,\rm {cm}^{-3}}$, ${\beta_0 = -2\times 10^3}$, and scalar field mass ${m_\varphi\, [1] =  10}$. The order $n$ defines the excitation mode of the scalar field solution to the eigenvalue equation \eqref{eq:perturbationequation_scalarfield}. The field $\varphi$ is shown in Einstein frame. The Jordan frame radius $r$ is normalized to the corresponding stellar radius $R$ for each solution.
    }
    \label{fig:scalarfield_profiles}
\end{figure}

The massless DEF model is strongly constrained by binary-pulsar observations, which require approximately $\beta_0\gtrsim-4.5$ \cite{Freire:2012mg,Antoniadis:2013pzd}, thereby fully excluding the possibility of white-dwarf scalarization. A finite scalar mass, however, Yukawa-suppresses the interaction on binary scales, substantially relaxing these bounds for $m_\varphi\gg10^{-16}\mathrm{eV}$ \cite{Ramazanoglu:2016kul}. Previous studies further restricted $|\beta_0|\lesssim10^3$ by requiring neutron stars, but not white dwarfs, to scalarize \cite{Popchev:2018fwu}; this is an assumption that we relax here. Conversely, scalarization inside a white dwarf requires a sufficiently light field, yielding $m_\varphi\lesssim10^{-11}\mathrm{eV}$ for $|\beta_0|\sim10^6$ and a compact Chandrasekhar-mass star. The same mass range is also compatible with cosmological evolution toward GR in massive scalar-tensor gravity \cite{dePireySaintAlby:2017lwc,Chiba:2021rqa}. Thus, a parameter region remains in which white-dwarf scalarization is compatible with both binary-pulsar and cosmological considerations.

\textit{Numerical results.} {We construct fully nonlinear equilibrium solutions. A summary of the theory parameter values which  were considered for numerical solutions are listed in Tab.\ \ref{table:numericalsolutions}. We specify values of $\beta_0$ and the dimensionless scalar field mass
 \begin{equation}
    m_{{\varphi}}[1]=\frac{m_{{\varphi}}}{m_{{\varphi},\mathrm{dim}}} \quad\mathrm{with}\quad m_{{\varphi}, \mathrm{dim}} = \frac{{c\hbar}}{10^7M_\odot\; G}\;,
     \label{eq:scalarfieldmassdim}
\end{equation}
such that $m_\varphi\, [1] = 1$ is equivalent to $1.3 \times 10^{-17}\, \rm eV$. We state whether scalarized solutions were found. Note that we find scalarized solutions for values of $|\beta_0|$ lower than the lowest theoretical estimate $|\beta_{\rm crit}| = 360$. This is consistent with our expectations. Our theoretical estimate assumes an even density distribution, $\tilde\rho_0 = 3M/4\pi R^3$, meaning that the central density of the star is underestimated. Consequently, since the critical value $|\beta_{\rm crit}|$ is approximately proportional to $1/\tilde \rho_0$, this threshold is overestimated. Branches represent different excitation modes $n$ of the scalar field for the same set of theory parameters. The number of obtained branches corresponds to the quantity of different eigenmodes $\omega_n$ that solve the eigenvalue equation \eqref{eq:perturbationequation_scalarfield}. We find up to four branches, $n = 0, 1, 2, 3$, for a specific parameter set that diverge from GR before the neutron-drip ${\rho_{\rm drip}\simeq 4.3\times10^{11}\, \rm g\, \rm cm^{-3}}$.}

{The excitation modes $n$ can be clearly seen in Fig.\ \ref{fig:scalarfield_profiles}. We show the radial scalar field solutions obtained for a star with central density ${\rho_c = 7.7\times10^{10} \, \rm g \, \rm {cm}^{-3}}$ as well as theory parameters ${\beta_0 = -2\times 10^3}$ and ${m_\varphi\, [1] =  10}$. Three excitation modes are present for this parameter set. The lowest branch, $n=0$, has no zero-crossing points. The following excitation modes, ${n=1, 2}$, have one and two zero-crossing points respectively. {All solutions are Yukawa suppressed by the scalar field mass and converge toward zero at infinity, ensuring flat asymptotics.}}

\begin{figure*}[p]
    \centering

    % Row 1
    \includegraphics[width=0.47\textwidth]{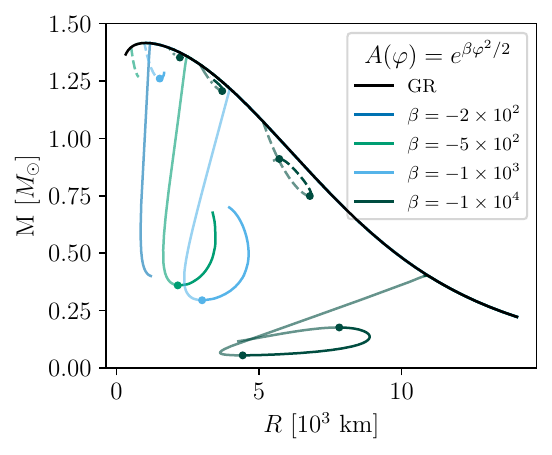}
    \hfill
    \includegraphics[width=0.47\textwidth]{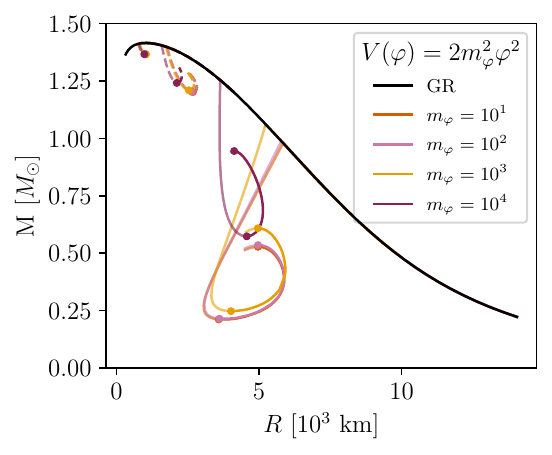}

    \vspace{0.5em}

    % Row 2
    \includegraphics[width=0.47\textwidth]{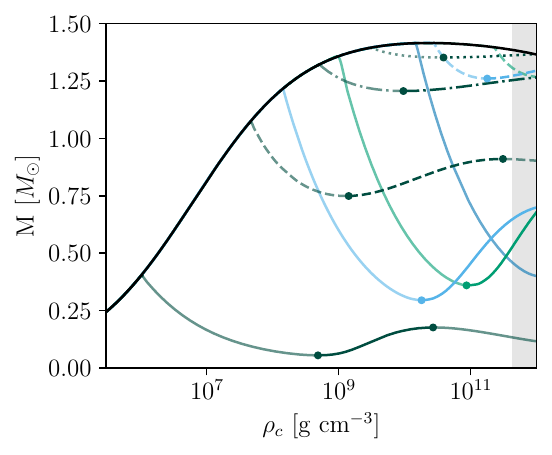}
    \hfill
    \includegraphics[width=0.47\textwidth]{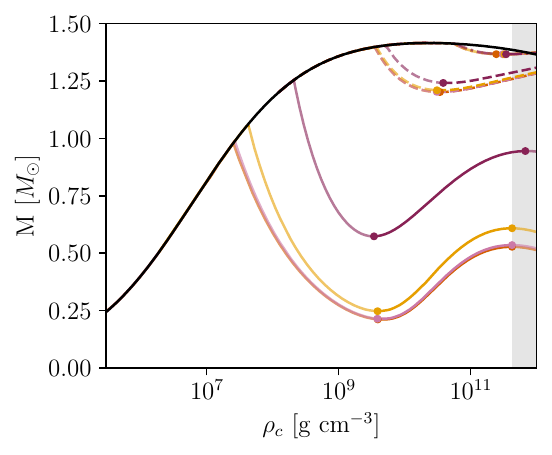}

    \vspace{0.5em}

    % Row 3
    \includegraphics[width=0.47\textwidth]{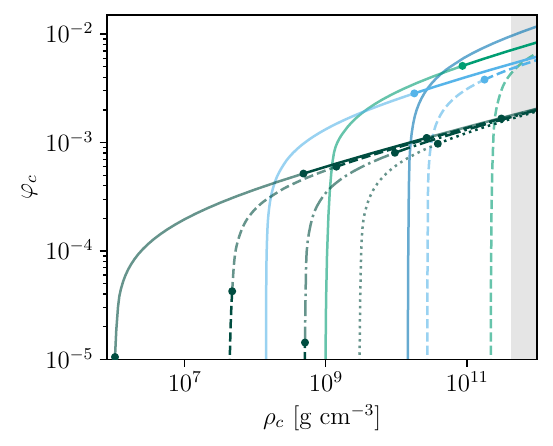}
    \hfill
    \includegraphics[width=0.47\textwidth]{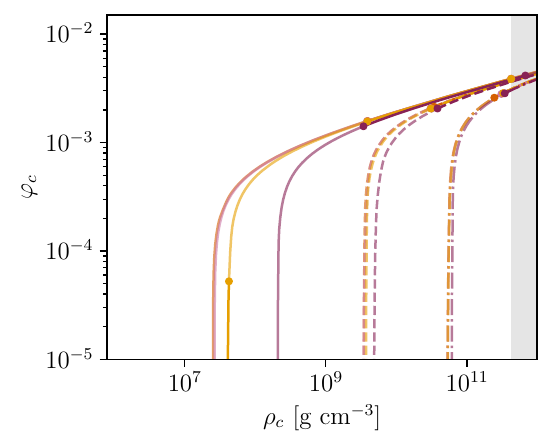}
    \caption{
        Numerically obtained scalarized solutions for a massive DEF model. Plots on the LHS show results for $m_\varphi [1]= 10$ and different values of $\beta_0$. On the RHS $\beta_0 = -2 \times10^3$ is fixed, while $m_\varphi$ is varied. We illustrate the resulting mass-radius (top), mass-density (middle), and central scalar field $\varphi_c$-density relations (bottom panels). Linestyles represent branches for $n=0$ (solid), $n=1$ (dashed), $n=2$ (dash-dotted), and $n=3$ (dotted) corresponding to different excitation modes of the scalar field solution. Local extrema of $M(\rho_c)$ are represented by filled circles in all plots. Transparent lines denote segments with $dM/d\rho_c<0$. Grey shaded regions indicate densities above the neutron-drip $\rho_{\rm drip}\simeq 4.3\times10^{11}$ g cm$^{-3}$.
    }
    \label{fig:comparison}
\end{figure*}

{Representative results for ${|\beta_0| \in [2 \times 10^2, 10^4]}$ and scalar-field masses ${m_\varphi [1] \in [10, 10^4]}$ are shown in Fig.\ \ref{fig:comparison}. The top panels display the numerically obtained mass-radius relations. While at large stellar radii $R$ solutions coincide with GR, scalarization leads to a departure from GR toward smaller radii. The new branches result in a lower Chandrasekhar limiting mass as well as considerably denser white-dwarf configurations. The same effect can be recognized in the mass-density relation shown in the middle panels. At low central densities, the solutions closely follow the GR white-dwarf sequence. Once the central density becomes sufficiently high, however, the scalarization threshold is surpassed and a qualitatively new structure emerges: after reaching a local maximum, the mass decreases with increasing central density, passes through a second turning point, and forms a new ascending family with $dM/d\rho_c>0$. Configurations on this sequence retain WD-scale masses and radii of a few $10^3\,\mathrm{km}$, while reaching central densities of order ${10^{11}-10^{12}\, \rm g\, \rm cm^{-3}}$.
Transparent lines in Fig.\ \ref{fig:comparison} denote segments with \(dM/d\rho_c<0\), suggesting configurations that are unstable against radial perturbations, as small deviations from equilibrium may lead to gravitational collapse or expansion rather than restoring the star to its original state.}
%\aw{another word we could use is sub-branches for omegas if family/sequence is too wide  etc}

{The lowest panels, illustrating the scalar field-density relation, underline that these are indeed scalarized solutions, i.e.\ configurations with $\varphi \neq 0$. A higher excitation $n$ of the field is reached for lower values of $\varphi_c$ at the same central density. As seen on the LHS panel, a more negative $\beta_0$ leads to an onset of scalarized solutions at higher central densities. We also observe that for larger $|\beta_0|$, the maximal scalar field value reached is higher. These effects are consistent with the theoretical expectations from the eigenmode equation \eqref{eq:perturbationequation_scalarfield}. Additional branches, $n=1, 2, 3$, set in at higher densities, while the value of the field converges toward the same value at higher central densities. On the RHS panel, we see that a higher scalar field mass also pushes the onset of the scalarized branch toward a higher central density, but does not affect the highest field value reached.}

{\textit{Discussion.}}
Remarkably, the phenomenon observed here resembles that of strange dwarfs \cite{Glendenning:1994zb,Vartanyan:2004dy} and, more generally, {third-family} compact stars: an additional sequence of equilibrium configurations is separated from the ordinary one by a decreasing-mass segment. The underlying mechanism is fundamentally different. Strange-dwarf configurations arise from exotic high-density matter, whereas here no additional matter component or phase transition is introduced. The same Chandrasekhar EoS is used throughout, and the additional family is generated entirely by the nonlinear development of the scalar field. Scalarization can therefore lead to the emergence of additional branches of equilibrium configurations, although the resulting mass–radius relations can differ significantly.

Some configurations reach densities close to or above the neutron-drip density, $\rho_{\rm drip}\simeq4.3\times10^{11}$ g cm$^{-3}$, where the Chandrasekhar EoS is no longer quantitatively reliable. Importantly, the first additional ascending branch already emerges below this density for different parameter sets, indicating that its onset is not caused by extending the EoS beyond the neutron-drip regime. A realistic high-density EoS will be required to follow the branch further and determine its quantitative properties.

\textit{Conclusions.} We have shown that spontaneous scalarization can generate a qualitatively new family of white-dwarf-like equilibrium configurations without invoking exotic matter or a phase transition. In massive DEF gravity, scalarized dwarfs remain possible within a parameter region compatible with existing binary-pulsar and cosmological considerations. Once sufficiently strong, the scalar field profoundly modifies the stellar sequence. Firstly, the Chandrasekhar limiting mass is slightly to significantly lowered depending on the theory parameters. Further, after the ordinary white-dwarf sequence reaches a turning point, the mass decreases and subsequently rises again, producing an additional sequence of considerably more compact configurations. Remarkably, this phenomenon resembles that of strange dwarfs and third-family compact stars \cite{Glendenning:1994zb,Vartanyan:2004dy}, although here it is generated entirely by the gravitational sector while the microscopic matter EoS remains unchanged.

The new family already emerges below the neutron-drip density, where the Chandrasekhar EoS employed here remains applicable, although following it to higher densities will require a realistic treatment of the matter composition. Whether these configurations constitute dynamically stable branches remains an open question and requires a radial-perturbation analysis including the scalar degree of freedom.

The possibility of such objects is particularly intriguing in view of the growing observational evidence for compact objects departing from standard mass-radius expectations. Candidate strange dwarfs have already been identified among sources classified as white dwarfs \cite{Kurban:2020xtb}, while objects such as the unusually low-mass compact star in HESS~J1731-347 \cite{Doroshenko:2022nwp}  illustrate that the observed compact-star population may contain configurations difficult to accommodate within conventional classifications. Neither provides a direct counterpart of the scalarized branches found here, but they motivate searches beyond standard white-dwarf and neutron-star sequences. Independent measurements of mass and radius in such systems could therefore provide a direct test of scalarization in a previously unexplored compactness regime.

\section*{Acknowledgements}
The authors acknowledge the support provided by the COST Action FuSe (CA24101). LJ was supported by the Estonian Research Council team grant ``Space - Time - Matter'' (PRG2608) as well as Estonian Ministry of Education and Research Centre of Excellence ``Foundations of the Universe'' (TK202U4). We would like to thank Daniela Doneva and David Alvarez Castillo for enlightening discussions.

{\section*{Data Availability Statement}}

{The data underlying the figures and numerical results presented in this work are available from the corresponding author upon reasonable request.}

\bibliographystyle{utphys}
\bibliography{biblio}

\end{document}